# Discovery of unconventional chiral charge order in kagome superconductor $KV_3Sb_5$

**Authors:** Yu-Xiao Jiang[1]*, Jia-Xin Yin[1]*†, M. Michael Denner[2]*, Nana Shumiya[1]*, Brenden R. Ortiz[3]*, Gang Xu[4]*, Zurab Guguchia[5], Junyi He[4], Md Shafayat Hossain[1], Xiaoxiong Liu[2], Jacob Ruff[6], Linus Kautzsch[6], Songtian S. Zhang[1], Guoqing Chang[7], Ilya Belopolski[1], Qi Zhang[1], Tyler A. Cochran[1], Daniel Multer[1], Maksim Litskevich[1], Zi-Jia Cheng[1], Xian P. Yang[1], Ziqiang Wang[8], Ronny Thomale[9], Titus Neupert[2], Stephen D. Wilson[3], M. Zahid Hasan[1,10,11,12]†

**Affiliations:**

[1]Laboratory for Topological Quantum Matter and Advanced Spectroscopy (B7), Department of Physics, Princeton University, Princeton, New Jersey, USA.

[2]Department of Physics, University of Zurich, Winterthurerstrasse, Zurich, Switzerland.

[3]Materials Department and California Nanosystems Institute, University of California Santa Barbara, Santa Barbara, California, USA.

[4]Wuhan National High Magnetic Field Center & School of Physics, Huazhong University of Science and Technology, Wuhan, China.

[5]Laboratory for Muon Spin Spectroscopy, Paul Scherrer Institute, CH-5232 Villigen PSI, Switzerland.

[6]Cornell High Energy Synchrotron Source, Cornell University, Ithaca, New York, USA.

[7]Division of Physics and Applied Physics, School of Physical and Mathematical Sciences, Nanyang Technological University, Singapore.

[8]Department of Physics, Boston College, Chestnut Hill, Massachusetts, USA.

[9]Institut für Theoretische Physik und Astrophysik, Julius-Maximilians-Universität Würzburg, Am Hubland, Würzburg, Germany.

[10]Lawrence Berkeley National Laboratory, Berkeley, California, USA.

[11]Princeton Institute for the Science and Technology of Materials, Princeton University, Princeton, NJ, USA

[12]Quantum Science Center, Oak Ridge, TN, USA

†Corresponding authors, E-mail:

jiaxiny@princeton.edu; mzhasan@princeton.edu
*These authors contributed equally to this work.

**Intertwining quantum order and nontrivial topology is at the frontier of condensed matter physics[1-4]. A charge density wave (CDW) like order with orbital currents has been proposed as a powerful resource for achieving the quantum anomalous Hall effect[5,6] in topological materials and for the hidden phase in cuprate high-temperature superconductors[7,8]. However, the experimental realization of such an order is challenging. Here we use high-resolution scanning tunnelling microscopy (STM) to discover an unconventional charge order in a kagome material $KV_3Sb_5$, with both a topological band structure and a superconducting ground state. Through both topography and spectroscopic imaging, we observe a robust 2×2 superlattice. Spectroscopically, an energy gap opens at the Fermi level, across which the 2×2 charge modulation exhibits an intensity reversal in real-space, signaling charge ordering. At impurity-pinning free region, the strength of intrinsic charge modulations further exhibits chiral anisotropy with unusual magnetic field response. Theoretical analysis of our experiments suggests a tantalizing unconventional chiral CDW in the frustrated kagome lattice, which can not only lead to large anomalous Hall effect with orbital magnetism, but also be a precursor of unconventional superconductivity.**

The interdependence of geometry, correlations, and topology is pivotal to many vexing questions of condensed matter. Advance in pushing this frontier forward directly contributes to our fundamental understanding of quantum matter and the application of quantum materials such as in quantum information science and energy relevant technology. Owing to the unusual lattice geometry, electrons in kagome lattice systems can experience a nontrivial Berry phase and strongly enhanced interactions, leading to various topological and many-body phenomena. Equipped with high spatial resolution, electronic detection, and magnetic field tunability, state-of-the-art STM has played a key role in discovering, elucidating, and manipulating Chern phases and many-body phenomena in kagome lattice systems[3,9-20]. However, the nature of collective ordering states, including CDW and superconductivity, remains elusive in the kagome lattice system. Recently, $KV_3Sb_5$ was discovered to be a kagome superconductor[21-23] with $T_C = 0.93K$. Theoretically, the normal state band structure of this material was identified to feature a $Z_2$ topological invariant[23]. Moreover, the material exhibits a transport and magnetic anomaly around 80K, which is speculated to be due to certain electronic order formation at low temperatures[21]. A giant anomalous Hall effect is observed[22] at low temperature, while the material does not exhibit magnetic order[21]. These pioneering observations and experimental mysteries suggest a striking yet unknown order that intertwines with nontrivial band topology, giant anomalous Hall effect and kagome superconductivity. Here we perform STM experiments on $KV_3Sb_5$ at 4.2K to observe a CDW order with an unusual magnetic field response.

$KV_3Sb_5$ has a layered structure with the stacking of $K_1$ hexagonal lattice, $Sb_2$ honeycomb lattice, $V_3Sb_1$ kagome lattice, and $Sb_2$ honeycomb lattice (Fig. 1**a**). Owing to the bonding length and geometry, the V and Sb layers can have stronger chemical bonding, and the material tends to cleave between K and Sb layers. According to the crystalline symmetry, at a monolayer atomic step edge between the K and Sb layers, the upper step will be K while the lower is the Sb layer (Fig. 1**b**). The step edge STM topography shown in Fig. 1**c** thus allows us to determine that the upper surface with vacancies is the K surface, and the lower surface with adatoms is the Sb surface, similar to the case[12,17] in the kagome magnet $Co_3Sn_2S_2$. The measured step height from topographic data is 2.3Å, which is close to the bulk structural distance between K layer and Sb layer as 2.23Å. Atomically resolved lattice topographies in the lower panels of Fig. 1**c** further confirm their respective hexagonal and honeycomb lattice structure. In additional, there are additional 2×2 superlattice modulations in the topographic data for both surfaces. Such modulations have not been observed in any other kagome lattice system[9-19], which can also have hexagonal and honeycomb surfaces. To study the nature of the 2×2 modulation, we focus on the Sb surface, which has a strong bonding with kagome lattice and features large defect-free areas. We measure these areas at 4.2K (Fig. 1**d**) and 80K (Fig. 1**d**), which is just above the critical temperature of the speculated electronic ordering[21]. It is clear that the 2×2 modulation disappears at 80K. Through the Fourier transform of the Sb topographic data, we further visualize the existence of 2×2 modulation vector peaks at 4.2K in Fig. 1**f** and its disappearance at 80K in Fig. 1**g**.

For an electronic order, the states in the vicinity of the Fermi level involved in its formation can produce an energy gap, such as in the classic Peierls mechanism for CDW order[24-27]. In our dI/dV data, measuring the local density of states, we observe a gap-like feature on the Sb surface in Fig. 2**a**. This energy gap extends from -23meV to +29meV, with additional shoulders around ±10meV, which could be due to the multi-orbital, multi-Fermi surface nature of the material[28] or anisotropy of the underlying order parameter. Moreover, we find this gap also disappears at 80K in our measurement, attesting to a close relationship with the observed 2×2 modulation. Our first-principles calculation reveals that the low-energy states are from V 3$d$ orbitals. Furthermore, the formation of V hexamers and trimers in the kagome lattice (as illustrated in the inset of Fig. 2**b**) with slightly reduced bond lengths than the original ones can reduce the total energy of the system by 32meV, rendering it a promising candidate for the 2×2 superlattice modulation. Our calculation of the local density of states (LDOS) of energy-optimized 2×2 superlattice structure also reveals an energy gap (Fig. 2**b**) around the Fermi level, which is of the same order of magnitude as the experimentally observed value. We further perturb the gap spectra by applying a magnetic field along the c-axis up to 6T, but we do not detect a strong field response of the gap structure (Fig. 2**c**). This observation is also consistent with a CDW gap.

For a CDW gap, it is also expected that across the energy gap, there should be an intensity reversal of the charge modulation[24-27]. In the classic Peierls CDW scenario, negative voltage bias shows enhanced intensity over charge accumulation regions, whereas images of the same atomic area at positive bias show enhanced intensity over charge depleted regions. The dI/dV imaging in Figs. 2**d** and **e** show that the maximum charge intensity at -30meV turns to the lowest charge intensity at +30meV despite the additional complexity of the modulation patterns. This observation thus demonstrates a type of charge intensity reversal. The charge modulation vector peaks are sharply evident in the Fourier transform of the spectroscopic imaging in Fig. 2**f**. This modulation vector (2×2) is non-dispersive within our q-resolution, as revealed by the energy-resolved Fourier transform of the spectroscopic imaging in Fig. 2**g**. The non-dispersive feature demonstrates a static electronic order. The topography and spectroscopic imaging taken together strongly support a CDW order.

A further inspection of our high-resolution charge modulation vector peaks in the low-energy spectroscopic data reveals pronounced intensity anisotropy along different directions, as shown in Fig. 3**a**. We take the data at surface defect-free region to study the intrinsic behavior of CDW, as defects, particularly those inducing standing waves, can backscatter electrons and pin (the phase of) the CDW order[24-26]. The observed anisotropy can be due to a chiral CDW order as observed and discussed in certain transition-metal dichalcogenides and high-temperature superconductors[29-32]. The chirality can be defined as the counting direction (clockwise or anticlockwise) from the lowest to highest vector peaks. In the TiSe$_2$ system, the chirality of the order can be manipulated by an optical field, which does not break time-reversal symmetry[30].

In the current case, we find the chirality at the same atomic area can be switched by the magnetic field applied along the c-axis for opposite directions, as shown in Figs. 3**b** and **c**. We have repeatedly observed the field switching effect at low-energies (relevant to the CDW gap) for different samples. Data taken at certain high energies do not show the switching effect. It is possible that this is due to additional contributions in this multi-orbital material, which deserves future study. The magnetic field switching effect suggests a time-reversal symmetry breaking of the CDW order, which is also supported by our recent muon spin spectroscopy measurement. Under similar magnetic fields, muon spin spectroscopy measurement reveals a sharp enhancement of magnetic response below the charge ordering temperature, which will be published elsewhere.

We try to understand the origin of this unconventional CDW order. The low energy band structure of $KV_3Sb_5$ consists of three nearly independent features (Fig. 4**a**): a quasi-two-dimensional electron pocket around the Γ point formed by $p_z$ orbitals from Sb; a band exhibiting a van Hove singularity at the M point, formed by the $d_{xy}$ orbitals of V; and a pair of Dirac-cone like bands near the M point, formed by $d_{xz}/d_{yz}$ orbitals of V. The band at van Hove filling stemming from V $d_{xz}/d_{yz}$ orbitals can be most vital to the formation of CDW order, since the vector of the 2×2 superlattice charge modulation connects the van Hove singularities at M points and matches with the Fermiology of the $d_{xz}/d_{yz}$ band[33-35]. Moreover, the kagome lattice at van Hove filling exhibits nested Fermi level eigenstates with unequal predominant sublattice occupancy[33] (Fig. 4**a**). Based on this sublattice interference mechanism[33], the predicted instability is a CDW order with relative angular momentum[35], which is consistent with a chiral charge order. Accordingly, we consider a triplet of chiral CDW order parameters[35] $\Delta_{Q1}$, $\Delta_{Q2}$, and $\Delta_{Q3}$, where $\Delta_{Qn}$, n = 1, 2, 3, are complex numbers whose absolute value approximately corresponds to the observed peak heights in Fig. 3 (see Fig. 4**a** for more details). The observed magnetic field response suggests that the time-reversal symmetry is broken by the CDW order parameters. This is typically achieved if degenerate order parameter components acquire a complex relative phase. Under reversal of time, such a phase changes sign, and thus, if it is not 0 or π, the state breaks time-reversal symmetry. One of the natural phase choices for the triplet order parameters is then $\arg(\Delta_{Qn}) = 2\pi n/3$. Applying this experimentally motivated choice of order parameters to a kagome model for the V $d_{xy}$ bands yields the energy gap and chiral charge ordering pattern as shown in Fig. 4**b**.

We further discuss the implications of the unconventional chiral CDW in light of the topological fermions and superconducting ground state of the system. As the Dirac bands of the V $d_{xz}/d_{yz}$ orbitals are also near M points, such an unconventional CDW order will open a topological energy gap[5,6] at the Dirac cones, thereby introducing significant Berry curvature (Figs. 4**c** and **d**). As a consequence, the system would exhibit an anomalous Hall effect. A $k_z$ integrated model estimation detailed in the Supplementary information gives rise to a large anomalous Hall conductivity of 310 $\Omega^{-1}\text{cm}^{-1}$, consistent with the reported intrinsic anomalous Hall conductivity[21]. Concurrently, the same Berry curvature field can lead to orbital magnetism[12,36], which is estimated to be 0.11$\mu_B$ per V atom, comparable with experimentally obtained value 0.22$\mu_B$ per V atom[21]. The orbital magnetization can couple to the external magnetic field, consistent with our experimental observation of the field response in Fig. 3. The magnetic field affects the phase of the order parameter through coupling with orbital magnetization, but not the amplitude of the order parameter. In reference to the saturation field of ±0.5T for the anomalous Hall conductivity[22], it would be interesting to explore the coercive field of the chirality switching effect in future experiments. Lastly, CDW and superconductivity are often intertwined with each other[1-4], as they are both collective phenomena that can be driven by similar interactions. The unconventional chiral CDW observed here can serve as a precursor for exotic Cooper pairing with finite relative angular momentum or broken time-reversal symmetry, providing a mechanism for unconventional superconductivity[1-4,35,37]. And it would be interesting to engineer the system down to atomic layers to look for other intimately related, yet hitherto unknown, topological quantum phenomena, including correlation-driven quantum anomalous Hall effect and chiral Majorana modes.

**Main figures**

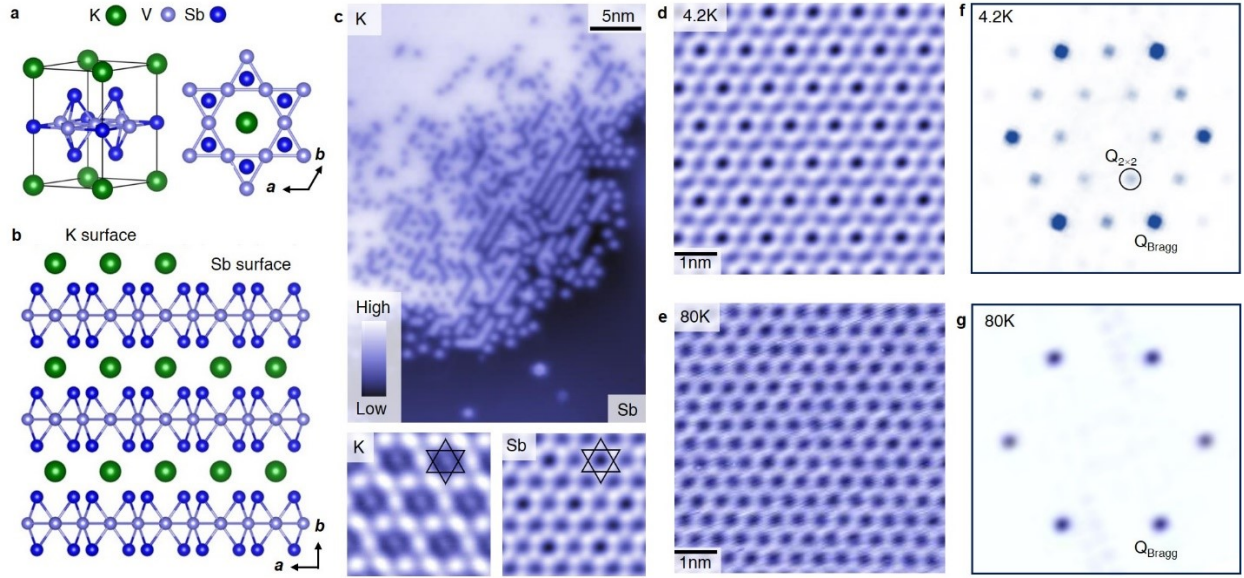

**Figure 1 Superlattice modulation observed via topographic imaging. a,** Crystal structure of $KV_3Sb_5$ from 3D view (left) and top view (right). **b,** Cleaving surfaces of $KV_3Sb_5$ illustrated from the side view of the crystal structure. **c,** Topographic image of a surface step edge, containing both K and Sb surfaces. The lower panels show atomically resolved topographic images of K hexagonal surface and Sb honeycomb surface, respectively. The black lines denote the underlying kagome lattice. **d,** A topographic image of a large Sb surface showing a 2×2 modulation. **e,** A topographic image of a large Sb surface taken at 80K showing absence of the 2×2 modulation. **f,** Fourier transform of the Sb topographic image, showing the ordering peaks and Bragg peaks. **g,** Fourier transform of the Sb topographic image taken at 80K, showing only Bragg peaks.

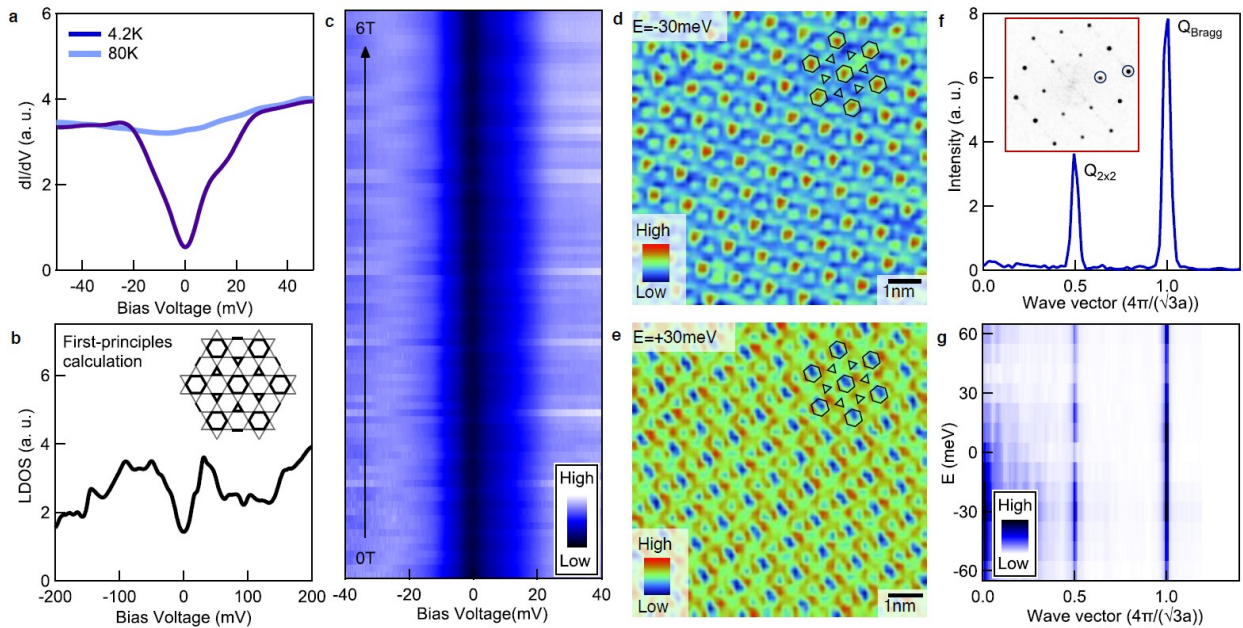

**Figure 2 Charge modulation observed via spectroscopic imaging. a,** Spatially averaged dI/dV spectrums for Sb surface taken at 4.2 and 80K, respectively. Each data is averaged over a defect-free area of 10nm×10nm. **b,** First-principles calculation of the bulk local density of states (LDOS) with considering the 2×2 superlattice modulation. The inset illustrates the charge order in the underlying V-based kagome lattice based on first-principles calculation, which forms hexamers and trimers (dark lines). **c,** Magnetic field perturbation of the dI/dV spectrums showing no detectable response. **d, e,** Atomically resolved dI/dV imaging for the same Sb surface at -30meV and +30meV, respectively. The inset shows the underlying ordered kagome lattice inferred from the simultaneously obtained topographic image. **f,** Fourier transform of the dI/dV imaging at -30meV (inset), showing both charge order and lattice Bragg peaks. **g,** Energy dependence of the charge order vector and lattice Bragg peaks, showing the non-dispersive nature of the charge order vector.

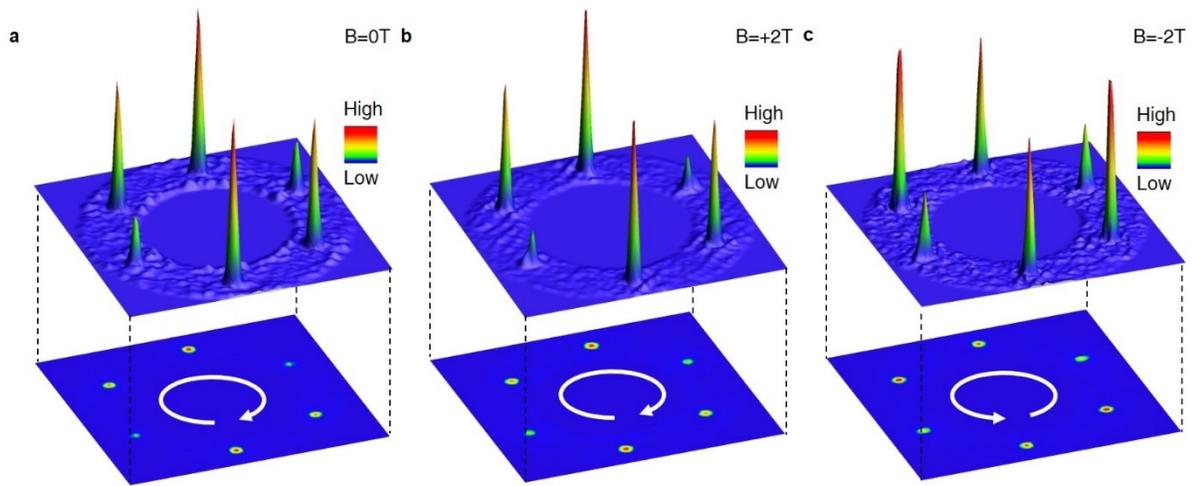

**Figure 3 Magnetic field response of the chiral charge order. a, b, c,** Spectroscopic 2×2 vector peaks taken at B=0T, -2T, +2T, respectively. Data are taken on defect-free regions. The images are Fourier transforms of spectroscopic maps acquired on an Sb surface 30nm×30nm in size at 10mV. A circular region of the full Fourier-transformed image is shown for clarify, highlighting the six 2×2 vector peaks. The top and bottom panels are 3D and 2D presentations of the data. The chirality can be defined as the counting direction (clockwise or anticlockwise) from the lowest to highest pair vector peaks.

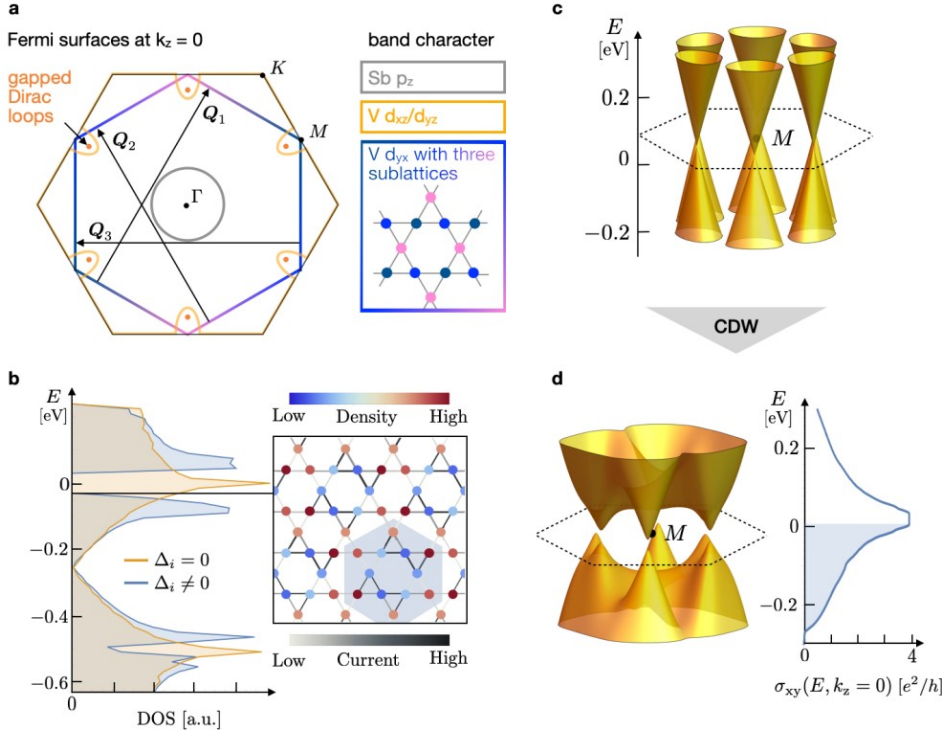

**Figure 4. Impact of unconventional charge order on the electronic structure. a,** Schematic of Fermi surfaces in the hexagonal Brillouin zone at $k_z = 0$. The wave vectors of the unconventional CDW are indicated, as well as the location at which Dirac nodal lines slightly above the Fermi energy cut the $k_z = 0$ plane. The V $d_{xy}$ Fermi surface, which is nested by the ordering wave vectors, has weight on distinct sublattices at each M point, giving rise to the sublattice interference mechanism. **b,** Two-dimensional model calculation of the impact of unconventional chiral CDW on the V $d_{xy}$ bands. The orange (blue) density of states is without (with) the CDW order parameter, which splits the van Hove singularity. Inset image shows the chiral charge pattern and associated orbital currents. The shaded area in the inset marks the 2×2 unit cell. **c,** Effective band structure of the nodal lines formed by the $d_{xz}/d_{yz}$ orbitals of V at $k_z = 0$ after 2×2 folding of the Brillouin zone. **d,** The introduction of a chiral time-reversal breaking CDW order parameter opens a topological gap around the Fermi-level. This gap gives rise to a non-zero Berry curvature, the integration of which produces a giant anomalous Hall effect for the bulk material.

## Methods

### Single-crystal growth

Single crystals of $KV_3Sb_5$ were synthesized from K (solid, Alfa 99.95%), V (powder, Sigma 99.9%) and Sb (shot, Alfa 99.999%). As-received vanadium powder was purified using EtOH and concentrated HCl to re-move residual oxides. Powder preparation was performed within an argon glove box with oxygen and moisture levels < 0.5 ppm. Single crystals of $KV_3Sb_5$ were then synthesized using a self-flux method using a eutectic mixture of $KSb_2$ and KSb, mixed with $VSb_2$. Elemental reagents were initially milled in a sealed, pre-seasoned tungsten carbide vial to form the precursor composition, which is approximately 50 at.% $K_xSb_y$ eutectic and approximately 50 at.% $VSb_2$. The precursor powder was subsequently loaded into alumina crucibles and sealed within stainless steel jackets. The samples were heated to 1000°C at 250 °C/hr and soaked there for 24 h. The mixture was subsequently cooled to 900°C at 100°C/hr and then further to 400°C at 2°C/hr. Once cooled, the flux boule is crushed, and crystals were extracted mechanically.

### Scanning tunneling microscopy

Single crystals with size up to 2mm×2mm were cleaved mechanically *in situ* at 77K in ultra-high vacuum conditions, and then immediately inserted into the microscope head, already at He4 base temperature (4.2K). More than 20 crystals were cleaved and studied in this research. For each cleaved crystal, we explore surface areas over 5μm× 5μm to search for atomic flat surfaces. Topographic images in this work were taken with tunneling junction set up V = 100mV I = 0.05nA for exploration of areas typically 400nm × 400nm. When we found atomic flat and defect-free areas, we took topographic images with tunneling junction set up V = 100mV I = 0.5nA to resolve the atomic lattice structure as demonstrated in the main paper. Tunneling conductance spectra were obtained with an Ir/Pt tip using standard lock-in amplifier techniques with a lock-in frequency of 997 Hz and a junction set up V = 50mV, I = 0.5nA, and a root mean square oscillation voltage of 0.3mV. Tunneling conductance maps were obtained with a junction set up V = 50mV, I = 0.3nA, and a root mean square oscillation voltage of 5mV. The magnetic field was applied with a zero-field cooling method. For field-dependent tunneling conductance spectra, we ramp the field continuously from 0T to 6T with 1T/hour ramp rate, with simultaneously compensating the field-induced spatial drift of the tip position on the sample. For the field-dependent tunneling conductance map, we first withdrew the tip away from the sample, and then slowly ramped the field to 2T or -2T. Then we reapproached the tip to the sample, found the same atomic area and then performed spectroscopic mapping at this magnetic field.

### Extended scanning tunneling microscopy characterization

Topographic images with a large field of view are shown in Fig. S1, where we found that the cleaving surfaces often consist of K clusters. Considering the charge activity of K ions, the possibility of surface charge polarization also deserves future study. There are also impurities inducing ring-like standing waves, which we try to avoid when study the intrinsic behavior of the charge order. An atom-resolved topographic image of the Sb honeycomb lattice is shown in Fig. S2 with a higher tunneling current I = 2nA. Figure S3 shows typical point spectrums for the Sb surface on three samples. We also performed dI/dV measurement of the K surface, which shows a slightly smaller energy gap as shown in Fig. S4. This difference could be due to certain orbital selectivity for tunneling with different surfaces[28].

Figure S5 shows extended dI/dV map data at different magnetic fields with different energies. From their Fourier transform we can determine the intensity anisotropy of 2×2 vector peaks. The magnetic field induced chirality switching is observed at low-energies relevant to the CDW gap.

Figure S6 shows the robust observation of strong intensity anisotropy of the 2×2 vector peaks for three samples with different scanning tip probes. We try to avoid the impurities with standing waves when taking the dI/dV maps to study the intrinsic behavior of the charge order.

Figure S7a shows the study at a defect-rich region, where several defects induce standing waves. The Fourier transform of the topographic data also reveals a ring-like signal within the 2×2 vector peaks (Fig. S7a inset). The defects backscatter electrons can strongly pin CDW order. Particularly, in q-space, the defects induced ring-like signal is close to the 2×2 vector peaks, implying a potential impact. In our dI/dV map data, we observe the chirality is strongly suppressed. Compared with the strong intensity variations for all three 2×2 vector peaks in Fig. S6, two vector peaks have similar intensity in Fig. S7b. In addition, the magnetic field response is also much weaker in this region, and there is no apparent chirality switch (Fig. S7b-d). We attribute these observations to the defects pinning effect of the CDW order.

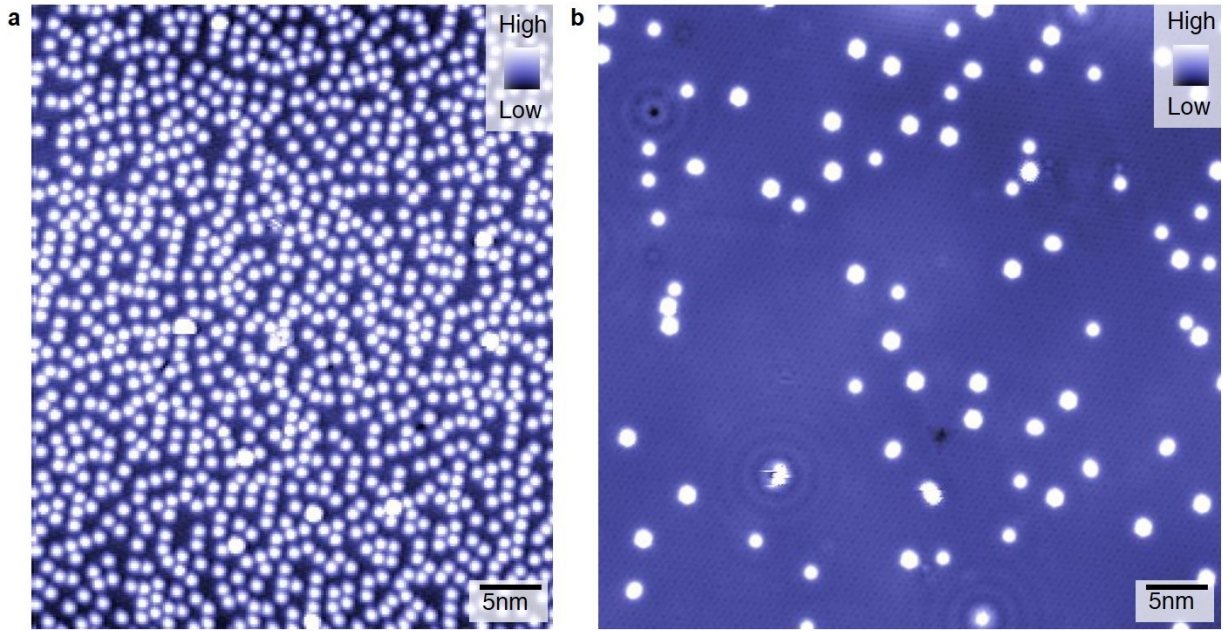

**Figure. S1 Typical topographic images over large areas. a,** Topographic image with more K clusters. **b,** Topographic image with fewer K clusters.

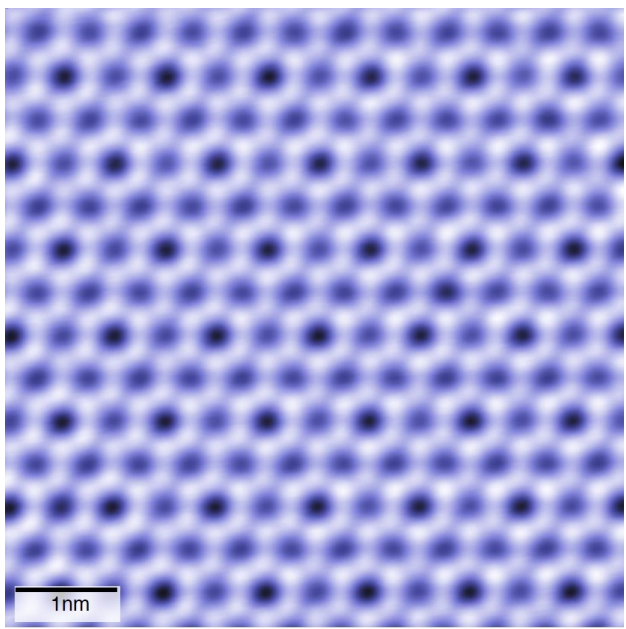

**Figure S2. An atom-resolved topographic image of the Sb honeycomb surface.**

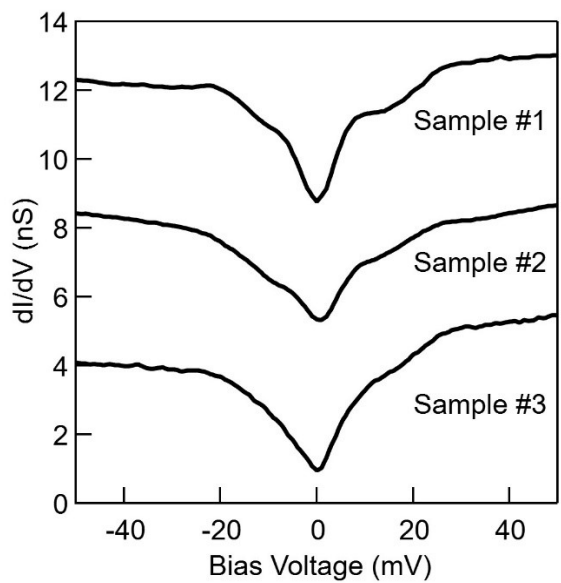

**Figure S3. Tunneling point spectrums taken on the Sb surface for different samples.** Spectrums are offset (with 4nS) for clarity.

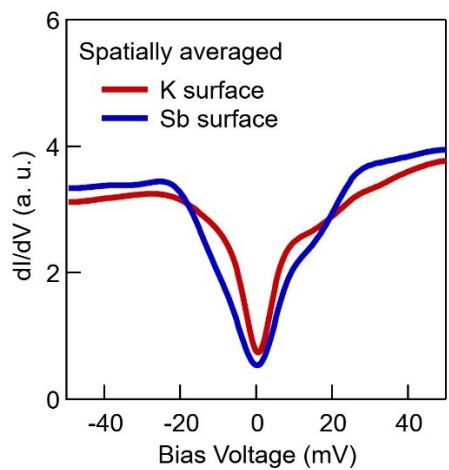

**Figure. S4 Comparison of the dI/dV spectrums taken at K surface and Sb surface.** Each data is averaged over a defect-free area of 10nm×10nm.

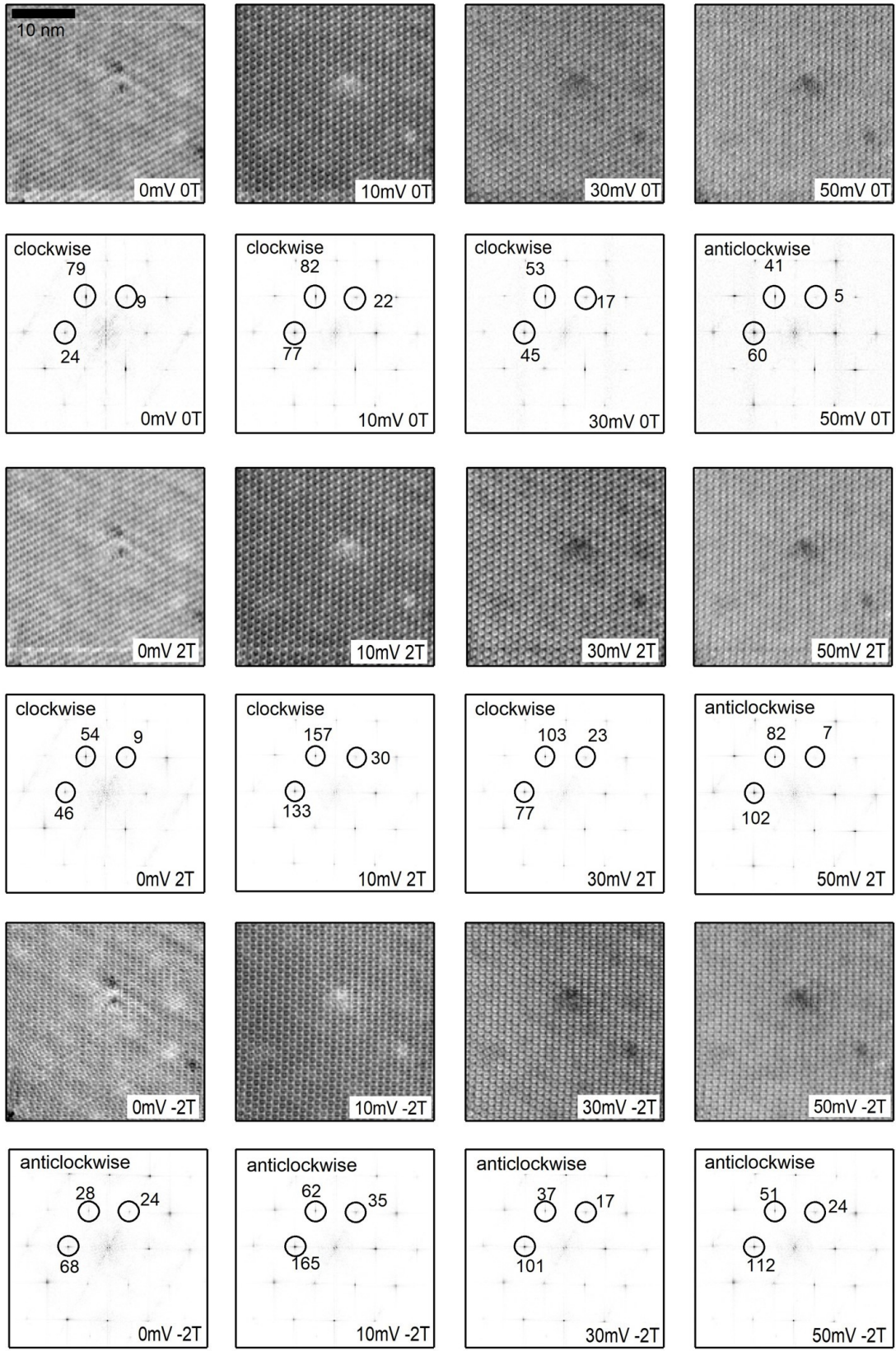

**Figure. S5 Spectroscopic maps and their Fourier transforms taken at different fields.** We observe that the chirality of low-energy charge modulations can be switched by the magnetic field. At 50meV, which is larger than the CDW energy gap, we do not observe the field switching effect.

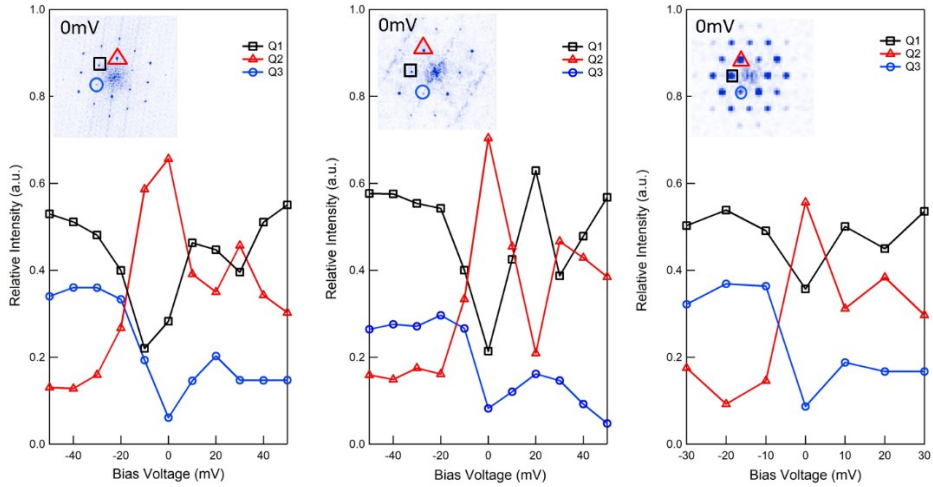

**Figure. S6 Robust observation of intensity anisotropy of 2×2 vector peaks at defect pinning free region.** We show the intensity of three CDW vector peaks as function of energy for three different samples. The maps are taken away from impurities with standing waves. The inset shows the Fourier transform of the zero-energy maps.

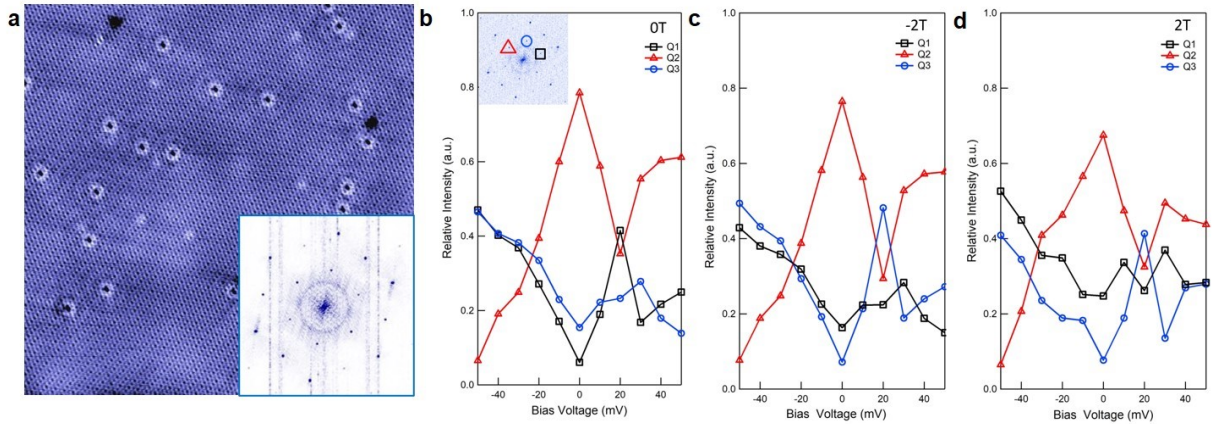

**Figure. S7 Defects pinning effect on charge order vector peaks. a,** Topographic data for a impurity rich region, where several defects induces standing waves. The inset is the Fourier transform of the topography, showing a ring-like signal corresponding to the standing waves. **b, c, d,** The intensity of three CDW vector peaks as function of energy for different magnetic fields. The data on this region shows the absence of clear chirality and absence of strong magnetic field response, indicating pinning effects of defects on CDW phase.

**First-principles calculations**

The low-energy electronic structure schematically summarized in Fig. 4a is based on the band structure depicted in Fig. S8. To obtain this band structure, we performed the density-functional theory calculations using the projected augmented wave method implemented in the Vienna ab initio simulation package[38-40] with generalized gradient approximation[41]. A k-mesh was used in the self-consistent computation. The cut-off energy for wave function expansion was set to be 400eV. The irreducible representation of energy bands was calculated using IrRep[42]. The 2×2 structures were optimized until the force on each atom was less than 0.001eV. The honeycomb Sb atoms show a similar bond-length modulation pattern as Fig. 1d, but the positions of K atoms and Sb atoms in the V-plane change slightly.

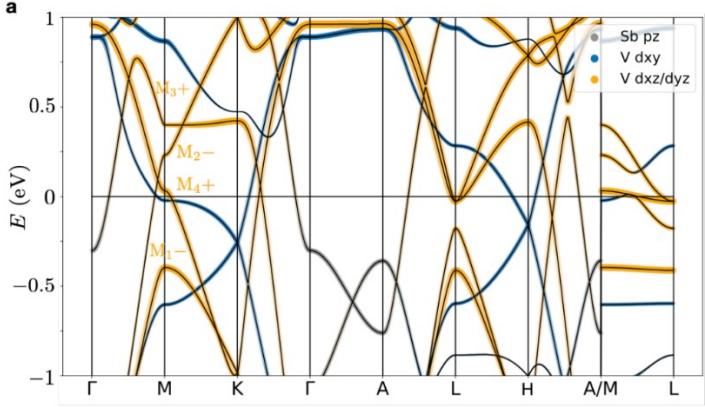

**Figure S8. Band structure of the hexagonal Brillouin zone without spin-orbital coupling.** The labels at the M point highlight the irreducible representations of the bands. Bands are colored according to their dominant orbital weight.

**Effective model for the charge order parameter**

To model the effect of the observed CDW on the low-energy electronic structure, we consider the kagome lattice formed by the $d_{xy}$ orbitals of V. At van Hove filling n= 5/12, the system exhibits a Fermi surface nesting. However, the Bloch states connected by the nesting wave vectors $Q_1$, $Q_2$, and $Q_3$ are localized on different sublattices. As introduced in Ref. 41, this sublattice interference mechanism leads to the formation of an unconventional CDW order, which modulates hopping amplitudes between the sublattices rather than onsite energies. As a result, the order parameter carries a nontrivial relative momentum structure. This can be described by an effective Hamiltonian of the form[41]:

$$H = \sum_{k,ij} H_{kagome;i,j}(k) c^\dagger_{k,i} c_{k,j} + \sum_{k,ijl} \Delta_{Q_j} \sin\left(\frac{Q_j k}{\pi}\right) |\epsilon_{ijl}| c^\dagger_{k,i} c_{k+Q_j,l} \quad (1)$$

where $H_{kagome}$ is the kagome lattice Bloch Hamiltonian of the $d_{xy}$ orbitals, the indices $i, j, l$, run over the three sublattices, and $\Delta_{Q_j}$ is the order parameter strength for the corresponding nesting vector $Q_j$. Concretely, we choose $\Delta_{Q_1} = 0.021$ eV, $\Delta_{Q_2} = 0.042$ eV $\omega$, and $\Delta_{Q_3} = 0.063$ eV $\omega^2$, with $\omega = e^{2\pi i/3}$, which have relative amplitudes that correspond to the experimentally observed anisotropic peak heights.

These non-zero order parameters gap the resulting band structure at the van-Hove filling and give rise to a chiral charge pattern as shown in Fig. 4b. Additionally, the complex phases of the order parameters spontaneously break time-reversal symmetry and give rise to non-zero currents, computed as $i\langle\psi|c^\dagger_i c_j - c^\dagger_j c_i|\psi\rangle$ shown as the bond strengths in Fig. 4b.

To incorporate the Dirac features, we rely on a k·p approximation around each of the three M points. We focus on the $d_{xz}/d_{yz}$ bands that form a Dirac-like band crossing slightly above the Fermi energy on the Γ-M line as shown in Fig. S8. At $k_z = 0$ and a given M point, the two bands involved are described by the Hamiltonian $H_M(k) = v_1 k_x \sigma_x + [v_2 k_y^2 - m^2]\sigma_z$, where $m/\sqrt{v_2}$ corresponds to the separation in momentum between the two Dirac points and $\sigma_{x/z}$ are the first and third Pauli matrix. The CDW folds all M points to the Γ point. The resulting effective Hamiltonian can then be written as[35,43]

$$H_C(\boldsymbol{k}_{x,y}) = \begin{pmatrix} H_M(\boldsymbol{k}_{x,y}) & \mathbb{1}_2 \Delta_{Q_1} & \mathbb{1}_2 \Delta_{Q_2} \\ \mathbb{1}_2 \Delta^*_{Q_1} & H_M(R_3 \boldsymbol{k}_{x,y}) & \mathbb{1}_2 \Delta_{Q_3} \\ \mathbb{1}_2 \Delta^*_{Q_2} & \mathbb{1}_2 \Delta^*_{Q_3} & H_M(R_3^2 \boldsymbol{k}_{x,y}) \end{pmatrix} \quad (2)$$

where $R_3$ is the 120° rotation matrix. The non-zero order parameters $\Delta_{Q_i}$ potentially gap the resulting band structure and equip the states with non-zero Berry curvature[44,45] when the phases of $\Delta_{Q_i}$ encode a chirality[46,47], spontaneously breaking time-reversal symmetry.

To estimate the magnitude of the Berry curvature-induced anomalous Hall conductivity that would result from the model in Eq. (1) we further introduce a $k_z$ dispersion which makes the model metallic[48]. Given the uncertainty about the form of the CDW order parameter, such an estimate can merely serve as a consistency check for the order of magnitude, rather than a quantitatively accurate point of reference. We calculate the intrinsic anomalous Hall conductivity through the integrated Berry curvature,

$$\sigma_{xy} = 2\frac{e^2}{h} \sum_n \int_{BZ} \frac{d^3k}{(2\pi)^3} \Omega_n(k), \quad (3)$$

where the sum runs over the occupied bands $n$ and $\Omega_n(k) = \partial_{k_x} i \langle u_{k,n} | \partial_{k_y} | u_{k,n} \rangle - \partial_{k_y} i \langle u_{k,n} | \partial_{k_x} | u_{k,n} \rangle$ is the Berry curvature of the corresponding band $|u_{k,n}\rangle$. To demonstrate that the correct order of magnitude (of about $\sigma_{xy} = e^2/h$ per layer of unit cell) can be obtained from our effective model, we choose the same order parameter values as above. We further choose $m^2 = 0.05$ eV, $v_1 = 0.3$ eV, $v_2 = 0.09$ eV, $\Delta = 0.07$ eV and add a $k_z$-dependent chemical potential $\tilde{\mu}(k_z) = (-0.03 \cos(k_z) + 0.02 \cos(2k_z) + 0.06)$ eV to the Hamiltonian in Eq(2). With this choice of parameters, one obtains a Hall conductivity (including spin degeneracy) of $\sigma_{xy} = 0.7 \frac{e^2}{h}$ per layer in z-direction, which amounts to 310 Ω⁻¹cm⁻¹. While the $k_z$ dispersion is quantitatively matched to the first principles calculation data, we emphasize again that this can only be an order of magnitude estimate due to the uncertainty in the CDW order parameter.

Same Berry curvature field will introduce orbital magnetism, which can be calculated by:

$$M = -i\frac{e}{2\hbar} \sum_n \int_{BZ} \frac{d^2k}{(2\pi)^2} f_{k_{x,y},n} \sum_{i,j=x,y} \epsilon_{ij} \langle \partial_{k_i} u_{k_{x,y},n} | [H_{CDW}(k_{x,y}) - E_{k_{x,y},n}] | \partial_{k_j} u_{k_{x,y},n} \rangle \quad (4)$$

with the Fermi-Dirac distribution $f$ as well as the energy $E$ of the eigenstate of $H_{CDW}$. We calculated the orbital moment to be $0.11 \mu_B$ per V atom.

**X-ray evidence for bulk CDW order**

High dynamic range x-ray diffraction maps (Fig. S9) were collected at the QM2 beamline at CHESS. The incident x-ray energy was 30 keV, selected using a double-bounce diamond monochromator. Temperature was controlled by bathing the small single crystal samples inside a stream of cold flowing helium gas. Diffraction was recorded in transmission though the sample using a 6 megapixel photon-counting pixel-array detector with a silicon sensor layer. Full 360 degree sample rotations, sliced into 0.1 degree frames, were indexed to the high-temperature crystal structure and transformed to reciprocal space. Some elements of the data reduction employed the NeXpy software package.

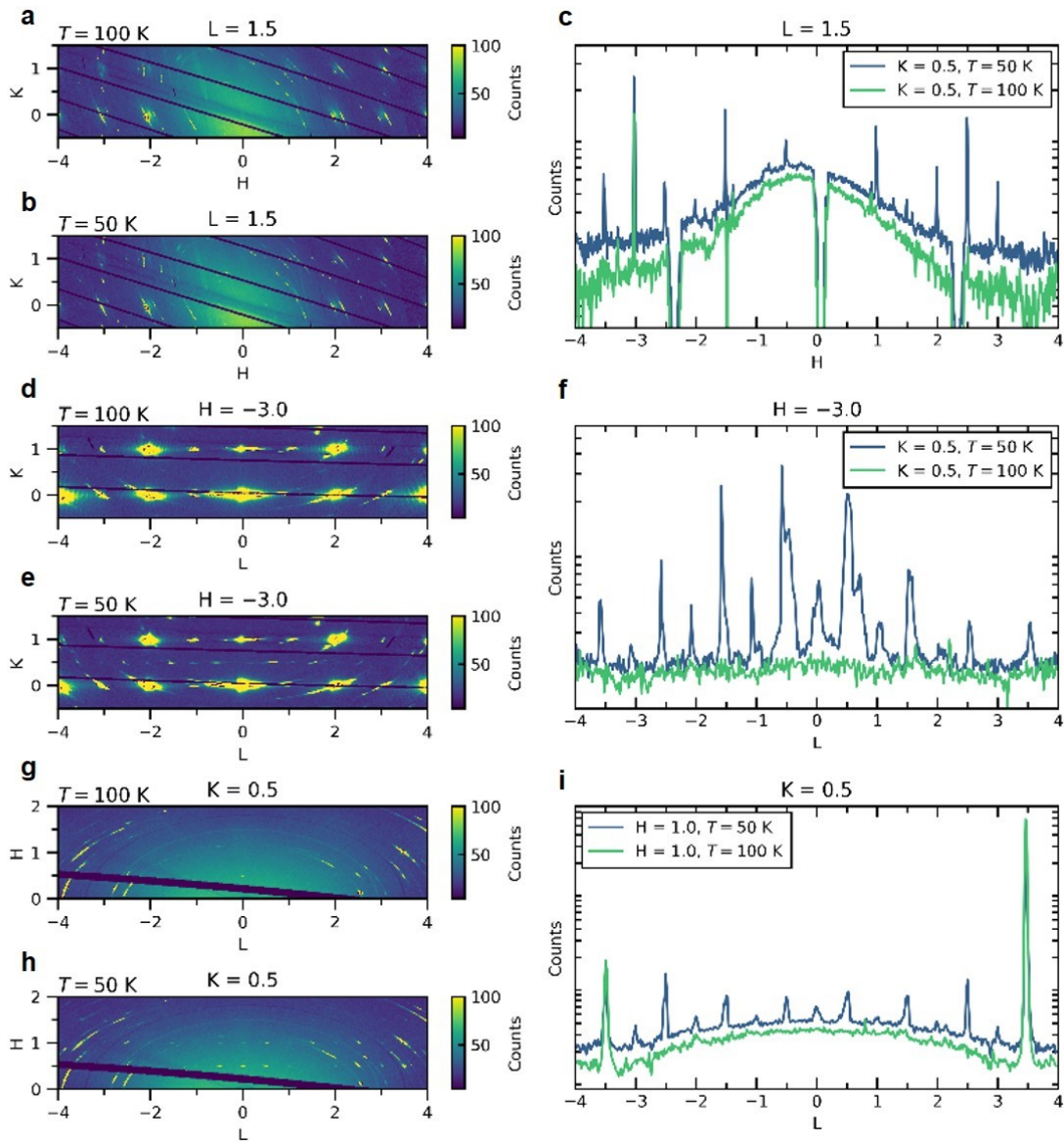

**Figure S9 Single-crystal synchrotron X-ray diffraction at T=100 K and T=50 K showing superlattice reflections at T=50 K with a propagation vector of (0.5, 0.5, 0.5) in HKL. a, b,** L=1.5 plane at **a** T=100 K and **b** T=50 K. **c,** Line cuts through the L=1.5 plane along the K=0.5 axis with changing H. **d, e,** H=-3.0 plane at **d** T=100 K and **e** T=50 K. **f,** Line cuts through the H=-3.0 plane along the K=0.5 axis with changing L. **g, h,** K=0.5 plane at **g** T=100 K and **h** T=50 K. **i,** Line cuts through the K=0.5 plane along the H=1.0 axis with changing L. The linecuts in **c, f, i** were created summing the over three neighboring axes to better capture the superlattice reflections. For example, the cuts along K=0.5 show the sum of the cuts along the axes K=0.49, K=0.50, and K=0.51.